\documentclass[twocolumn,showpacs,preprintnumbers,amsmath,amssymb]{revtex4}
\usepackage{colordvi,graphicx,color,amsbsy}

\usepackage{graphicx}
\usepackage{dcolumn}
\usepackage{bm}
\usepackage{enumerate}
\usepackage[latin1]{inputenc}
\newcommand{\comment}[1]{}

\begin{document}

\title{Lagrangian statistics and flow topology in forced two-dimensional turbulence}
\author{B. Kadoch \footnote{Current address:
Mechanical Engineering Program, University of San Diego, 5998 Alcala 
Park, San Diego, CA 92110, USA}}
\affiliation{M2P2-UMR 6181 CNRS \& CMI, \\ Universit\'e d'Aix-Marseille, Marseille, France}
\author{D. del-Castillo-Negrete}
\affiliation{Oak Ridge National Laboratory, \\ Oak Ridge, Tennessee, USA}
\author{W.~J.~T. Bos}
\affiliation{LMFA-UMR 5509 CNRS, Ecole Centrale de Lyon \\ Universit\'e Claude Bernard Lyon 1-INSA de Lyon, Ecully, France.}
\author{K. Schneider}
\affiliation{M2P2-UMR 6181 CNRS \& CMI, \\ Universit\'e d'Aix-Marseille, Marseille, France.}

\date{\today}
\pacs{47.27.E-, 47.27.T-, 47.27.N-}

\begin{abstract}
A study of the relationship between Lagrangian statistics  and flow topology in fluid turbulence 
is presented. The topology is characterized using the Weiss criterion that provides a simplified tool to partition the flow into topologically different regions: elliptic (vortex dominated), hyperbolic (deformation dominated), and intermediate (turbulent background).  The flow corresponds to forced two-dimensional Navier-Stokes turbulence in doubly periodic and circular bounded domains with non-slip boundary conditions.  In the double periodic domain, the probability density function (pdf) of the Weiss field exhibits a negative skewness consistent with the fact that in periodic domains the flow is dominated by coherent vortex structures. On the other hand, in the circular domain, the elliptic and hyperbolic regions seem to be statistically similar. We follow a Lagrangian approach and obtain the statistics by tracking  large ensembles of passively advected tracers. The pdfs of residence time in the topologically different regions are computed using the Lagrangian Weiss field, i.e., the Weiss field computed along the particles'  trajectories.  In elliptic and hyperbolic regions, the pdfs of the residence time have self-similar algebraic decaying tails. On the other hand, in the intermediate  regions the pdf has exponential decaying tails.  The conditional (with respect to the flow topology) pdfs of the Lagrangian velocity  exhibit Gaussian behavior in the periodic and in the bounded  domains. In contrast to the freely decaying turbulence case, the conditional pdfs of the Lagrangian acceleration in forced turbulence show a comparable level of intermittency in the periodic and the bounded domains. The conditional pdfs of the Lagrangian curvature are characterized, in all cases, by self-similar power law behavior  with a decay exponent of order $-2$.

\end{abstract}
\maketitle

\section{Introduction}

The emergence of coherent structures is a well-established phenomenon in two-dimensional turbulence \cite{McWilliams_1984}. These structures are characterized by localized regions of intense vorticity and have a direct impact on the flow dynamics and the transport \cite{Lesieur_fluids}. In particular,  vortices tend to trap tracers for long times, while strong shear can lead to large relative displacements of tracers \cite{Elhmaidi_JFM_1993,beto}. 
The combined effect of  vortex trapping and free streaming along shear flows is important because it can lead to anomalous diffusion \cite{del_castillo_1998,del_castillo_2005}.  Therefore, it is of significant theoretical and practical interest to understand the relation between coherent structures and transport. The present paper focuses on this problem from a Lagrangian perspective. 

By means of direct numerical simulation, we study transport of passive tracer particles in a turbulent flow. Due to the nontrivial spatio-temporal dynamics of the flow field, the particle trajectories are stochastic and one has to resort to statistical tools to characterize the Lagrangian dynamics. This leads to the problem of understanding the role of coherent structures on the Lagrangian statistics of tracers. Although important progress has been made, this problem is not well-understood. In particular, most studies have focused on the statistics of the Lagrangian displacements 
\cite{Elhmaidi_JFM_1993,Provenzale_CSF_1995,Castilla_NPG_2007}. Although this problem is of great interest for the understanding of dispersion and transport of passive scalars, there are other Lagrangian quantities of interest.
For example, the study of the statistics of the curvature and the acceleration of Lagrangian trajectories  in fluids has been the object of recent experimental \cite{xu_PRL_2007} and numerical studies \cite{Braun_JOT_2007,wilczek_PhysD_2008,Toschi_ARFM_2009}. 

The study of the acceleration and the curvature along particle trajectories in turbulent flows is directly related to the problem of intermittency and non-Gaussian statistics. This is because the Lagrangian acceleration and curvature time series typically contain ``bursty" phenomena which give rise to slowly decaying non-Gaussian tails in the corresponding pdfs.
In Ref.~\cite{xu_PRL_2007} a high speed particle tracking system was used to measure the curvature,
$\kappa= |a_n| u_L^{-2}$,
where $a_n$ is the normal acceleration and $\bm u_L$ is the Lagrangian velocity, of trajectories in a turbulent closed water flow between counter-rotating disks, and  it was shown that the probability density function (pdf) of the curvature values exhibits power law decay of the form $\sim \kappa^{-5/2}$. The same result was obtained in three-dimensional direct numerical simulations \cite{Braun_JOT_2007} . The statistics of the Lagrangian curvature and acceleration was also studied numerically in decaying two-dimensional turbulence in Ref.~\cite{wilczek_PhysD_2008} where it was shown that the Lagrangian acceleration exhibited exponential tails and the pdf of the curvature decayed as  $\sim \kappa^{-2.25}$. 

The curvature of fluid particle trajectories is directly related to the Lagrangian acceleration. This quantity has received considerable attention during the last decade, in particular in the framework of homogeneous and isotropic turbulence \cite{Yeung_ARFM_2002,Toschi_ARFM_2009}.   
 The effect of boundaries on the Lagrangian statistics was addressed in \cite{Kadoch_PRL_2008} where it was shown that, in the case of decaying turbulence, the Lagrangian acceleration is more intermittent  in confined domains than in periodic domains. 

\begin{figure}[!htb]
 \begin{center}
 \includegraphics[scale=0.4]{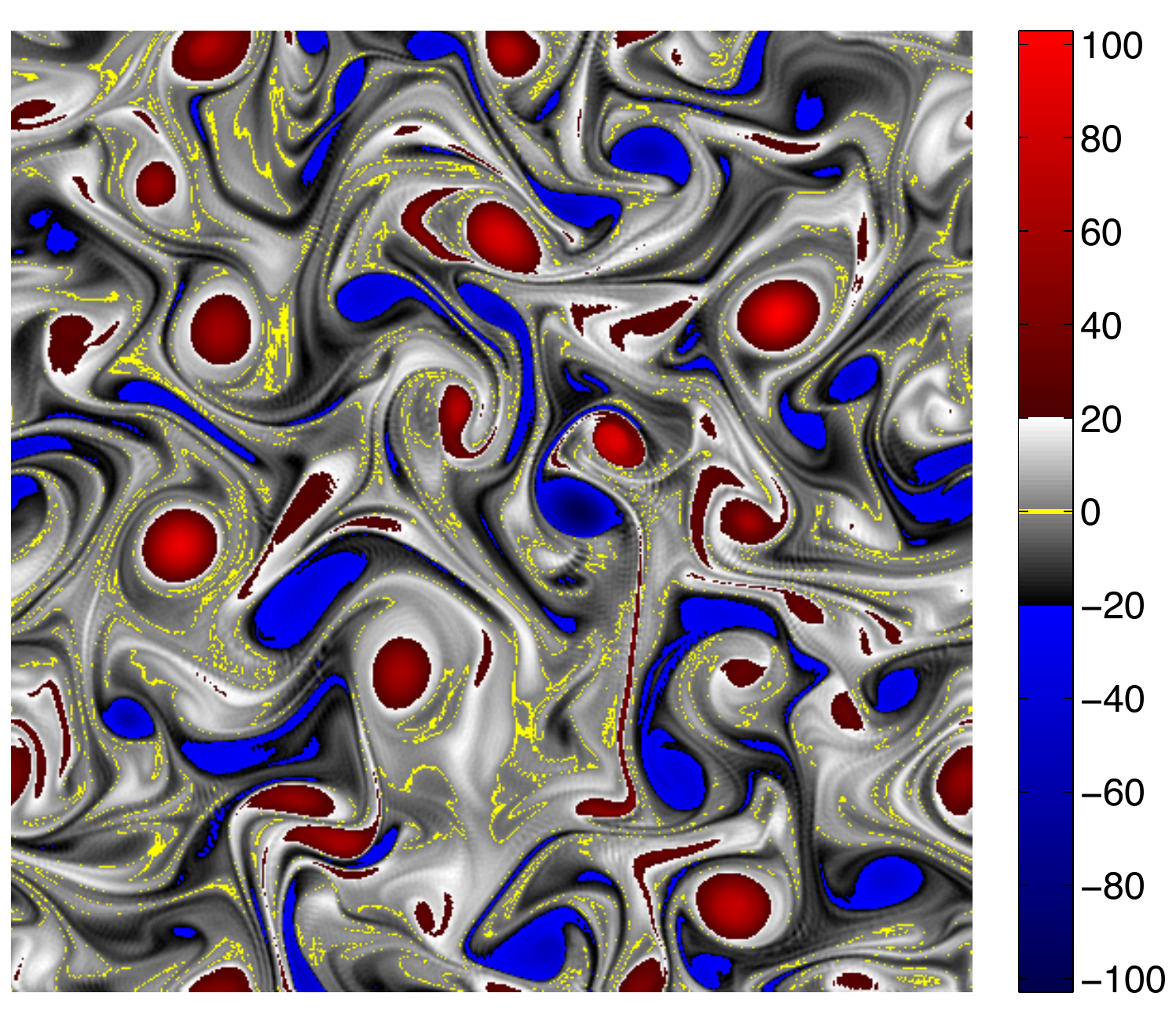}
 \includegraphics[scale=0.4]{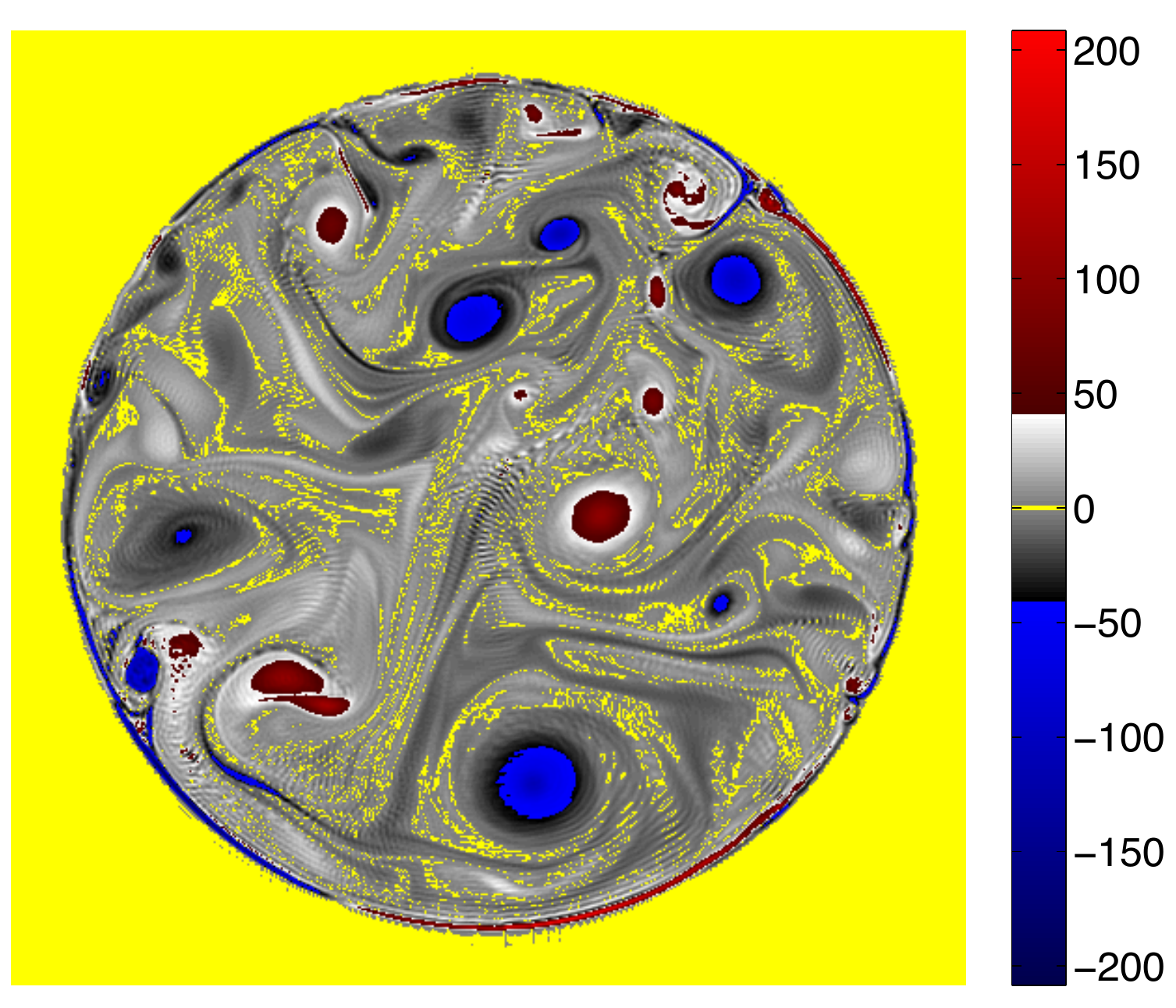}
\caption{Snapshots of vorticity field at t=25  in forced two-dimensional turbulence in a double periodic domain (top panel) and 
in a bounded circular domain (bottom panel).
Red denotes cyclones, $\omega >0$; Blue denotes denotes anti-cyclones, $\omega <0$; and Yellow denotes regions of vanishing vorticity.}
\label{fig_vorticity}
 \end{center}
\end{figure}

In this paper we focus on the influence of the flow-topology on the Lagrangian dynamics.
We investigate the statistics  of the Lagrangian velocity, acceleration and curvature, but in contrast to previous works, we study the problem of forced turbulence in unbounded and bounded domains. Most importantly, going beyond previous studies on the relation between the flow topology and Lagrangian statistics, we consider the  conditional probability  of the Lagrangian curvature and acceleration on the flow topology which we characterize using the Lagrangian Weiss criterion. 

The remainder of the paper is organized as follows. Section~II defines the forced, two-dimensional turbulence model in periodic and bounded circular domains, and discusses the numerical methods used in the integration of the model and in the integration of the Lagrangian orbits. Section~III discusses the characterization of the flow topology in terms of the Weiss field. The residence time statistics are presented in Section~IV. The conditional statistics (on the flow topology) of the Lagrangian velocity,  acceleration and curvature are discussed in Section~V. Section~VI presents the conclusions. 

\section{Turbulence model}

The turbulence model is based on the two-dimensional Navier-Stokes equations written in dimensionless form 
\begin{equation}\label{NS}
\frac{\partial \omega}{\partial t} + \bm V \cdot \nabla \omega - \nu \nabla ^2 \omega - F_\omega = - \frac{1}{\eta} \nabla \times (\chi  \bm V)\; ,
\end{equation}
\begin{equation}\label{NS2}
\nabla \cdot\bm V = 0,
\end{equation}
where, ${\bf V}=(u,v)$ is the flow velocity, $\omega =\nabla \times {\bf V}$ is the vorticity,  $\nu$ is the kinematic viscosity, and $F_\omega$ an external force. 
We consider two different types of domains: a double periodic domain (unbounded) and a circular domain with no-slip boundary conditions (bounded). The term on the right hand side of Eq.(\ref{NS}) is a volume penalization term which takes into account the no-slip boundary conditions using $\chi$ as mask function \cite{Angot_1999,Schneider_CF_2005,Schneider_PRL_2005}. This term is absent in the  periodic domain calculations. In the case of  bounded domains, the mask function vanishes inside the fluid domain ($\chi=0$), and is equal to one outside the fluid domain ($\chi=1$), where the no-slip boundary conditions are imposed. The permeability $\eta$ is chosen sufficiently small for given $\nu$  in order to insure the convergence of the volume penalization method \cite{Schneider_CF_2005}. Here we use $\nu=5 \times 10^{-4}$ and $\eta=10^{-4}$. For the periodic geometry, the Reynolds number is $Re_p=S\sqrt{E}/\nu = 2.5 \times10^{4}$, where $S=2\pi$ corresponds to the domain size and $E=1/2\langle {\bm V}^2 \rangle$ is the turbulent kinetic energy. For the circular geometry the Reynolds number is $Re_c=2r \sqrt{E}/\nu = 2.4 \times10^{4}$ where $r=2.8$ is the radius of the circle.
 
To obtain a statistically stationary flow we consider $F_{\omega} = F_{r} - \beta \psi$,  where $F_{r}$ denotes a random isotropic stirring at small ($k=8$) wave-numbers. The Rayleigh friction term, $- \beta \psi$, is added to avoid the accumulation of energy at large scales due to the inverse energy cascade. For the periodic domain we choose $\beta=1$. However, for the confined domain case we choose $\beta=0$ since the wall plays the role of an energy sink and there is no need for extra large-scale dissipation. In the confined domain, a time correlation is introduced into the forcing term. In the calculations presented here,  a discrete Markov chain is used as in Ref.~\cite{Lilly_1969},
$F(t_{n+1}) = \alpha F(t_n) + (1 - \alpha^2)^{1/2} F_{\omega}$
where $\alpha = \frac{1-\Delta t /\tau_c}{1+\Delta t /\tau_c}$
and  $\tau_c=10^{-2}$.
This time correlation is smaller than the viscous timescale, and only serves to improve the numerical stability. 

The numerical scheme is based on a classical pseudo-spectral method, fully dealiased, and the resolution is $N= 512^2$ grid points. The time-integration is of semi-implicit type with $\Delta t= 1.10^{-4}$ \cite{Schneider_CF_2005}. The particle trajectories were calculated by interpolating the Eulerian quantities and by using a second order Runge-Kutta scheme for time integration. The Lagrangian acceleration is the sum of the pressure gradient, viscous diffusion and external forcing 
\begin{equation}
\bm a_L = -\nabla p + \nu \nabla^2 \bm u + \bm f \, ,
\end{equation}
 where $\bm f$ is constructed by applying Biot-Savart operator to $F_{\omega}$. The Lagrangian statistics were computed by ensemble averaging over $10^4$ trajectories. Each trajectory was integrated for $5 \times 10^5$ time-steps, which correspond to 
 $800 \, \tau$, where $\tau=1/\sqrt{2 Z}= 0.0622$ is the eddy turnover time and 
$Z=1/2\langle \omega^2 \rangle$ the enstrophy.
Figure~\ref{fig_vorticity} shows snapshots of the vorticity field at a fixed time in the periodic and the bounded domains. The characteristic
emergence of coherent vortices in two-dimensional turbulence can be observed. Note the relatively strong vorticity values at the non-slip boundary  in the  circular domain. 

\section{Flow topology}

Since the flow is two-dimensional and incompressible, ${\bf V}=\hat{z} \times \nabla \psi$, where $\psi(x,y,t)$ is the streamfunction. The Lagrangian orbits are then obtained from the solution of the Hamiltonian system
\begin{equation}
\label{hamilton}
\frac{d x}{dt}= -\frac{\partial \psi}{\partial y}\, , \qquad \frac{d y}{dt}= \frac{\partial \psi}{\partial x}\, , 
\end{equation}
where $\psi$ plays the role of the Hamiltonian, and $(x,y)$ are the canonically conjugate variables. 
By flow topology we mean the topology of the streamlines, which  corresponds to the topology of the Hamiltonian. The study of the location and dynamics of the fixed points of the Hamiltonian, which are the stagnation points of the flow, is one of the most natural  topological consideration with direct impact to transport. Elliptic fixed points give rise to trapping regions and hyperbolic fixed points induce rapid mixing and stretching \cite{ottino}. However, the nontrivial spatio-temporal dependence of fluid turbulence makes the identification and tracking  of the elliptic and hyperbolic regions rather difficult. Among the experimental works addressing this problem are Refs.~\cite{gollub_PRL_2007,gollub_PoF_2008} which proposed a method to  characterize the elliptic and hyperbolic points of flows exhibiting spatio-temporal chaos  based on the  measured Lagrangian curvature of tracer-particle trajectories. These, as well as other experimental studies (see for example the recent review in \cite{Toschi_ARFM_2009}), have pointed out the need for 
a deeper understanding of the impact of the flow topology on transport and the need for novel diagnostics to understand the Lagrangian statistics.

The characterization of the flow topology used here is based on the Okubo-Weiss criterion \cite{Okubo_1970,Weiss_PysD_1991} which is an approximate method of partitioning a flow field into topologically distinct regions. From the physics point of  view the flow is divided into  vorticity dominated regions, which correspond to elliptic regions, strong deformation regions which correspond to hyperbolic regions and intermediate regions. The partition is based on the relative value of 
\begin{equation}
\label{weiss}
Q=s^2-\omega^2 \, ,
\end{equation}
where $\omega= \partial_x v-\partial_y u$ is the vorticity and $s^2=s_1^2+s_2^2$ is the deformation where 
 \begin{equation}
 s_1=\frac{\partial u}{\partial x}-\frac{\partial v}{\partial y} \, , \qquad 
 s_2=\frac{\partial v}{\partial x}+\frac{\partial  u}{\partial y} \, .
\end{equation}
Using the local value of $Q$, the flow domain can be partitioned in three disjoint regions:
\begin{itemize}
\item[i)]  strongly elliptic for which $Q \leq -Q_0$
\item[ii)] strongly hyperbolic for which $Q \geq Q_0$
\item[iii)] intermediate regions for which $-Q_0<Q<Q_0$
\end{itemize}
where $Q_0$ is the standard deviation of the values of  $Q$, i.e., $Q_0=\sqrt{\langle Q^2 \rangle}$ where $\langle \cdot \rangle$ is the ensemble average. Although this way of partitioning the flow has limitations, since it is based on the linearization of the Navier-Stokes equation \cite{Basdevant_PysD_1994}, it is a conceptually simple tool that has provided key insights in the understanding of the dynamics of turbulence and Lagrangian transport in fluids and plasmas \cite{Elhmaidi_JFM_1993,Annibaldi_2002}.

\begin{figure}[!htb]
 \begin{center}
 \includegraphics[scale=0.4]{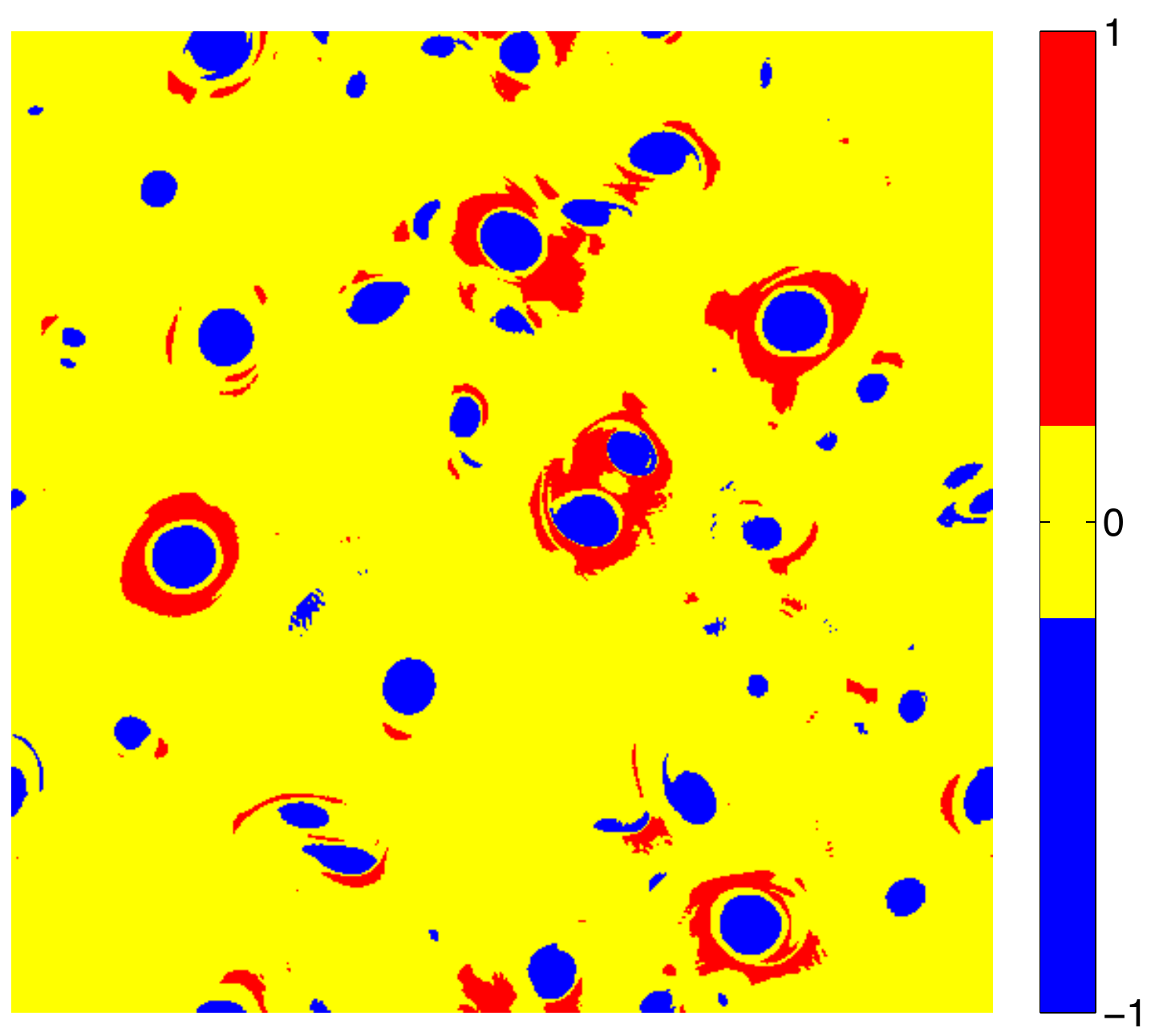}
 \includegraphics[scale=0.4]{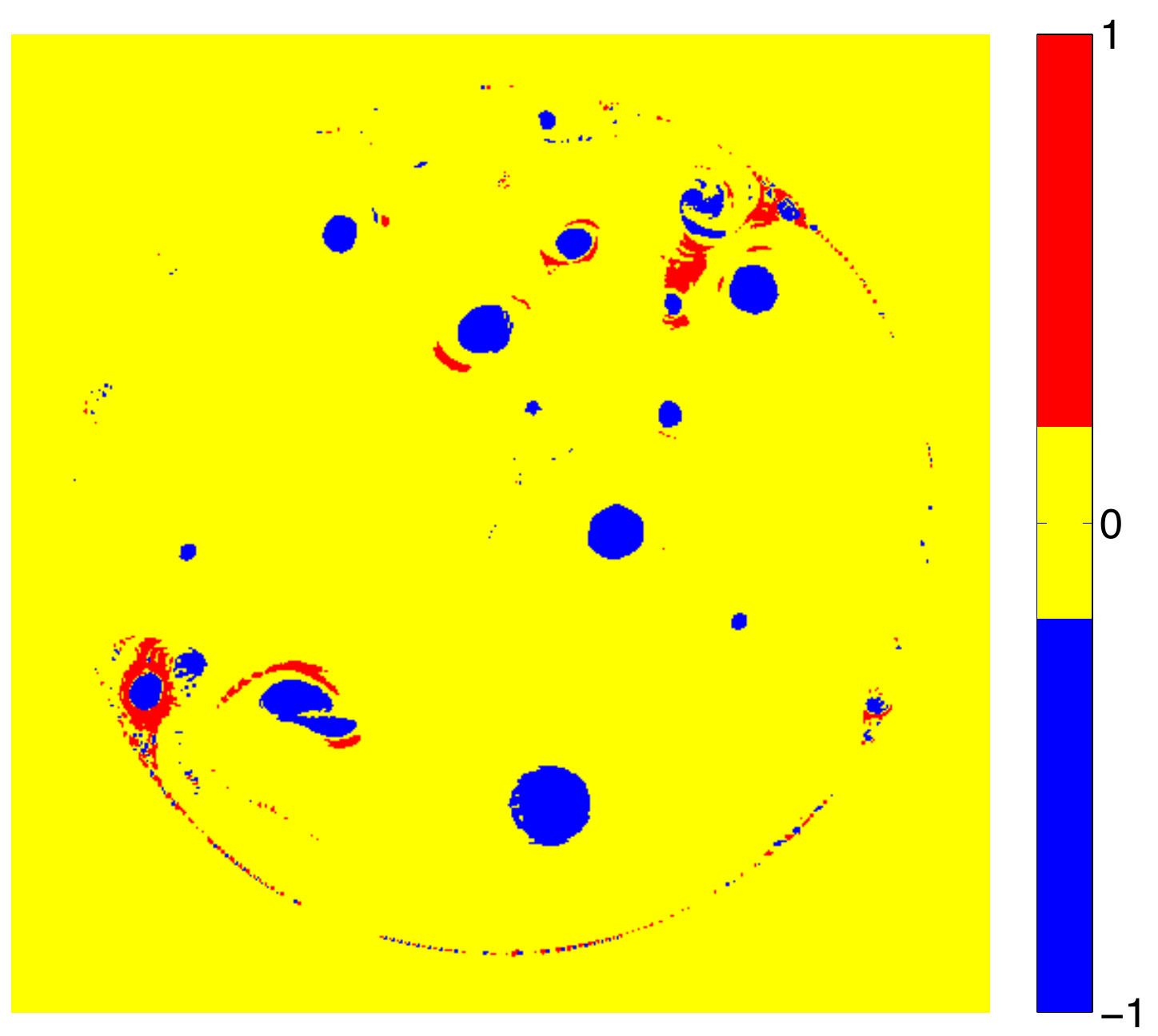}
\caption{Snapshots of three levels Weiss field, $\hat Q=Q/Q_0$, at t=25  in forced two-dimensional turbulence in a double periodic domain (top panel) and in a bounded circular domain (bottom panel). The flow fields correspond to those shown in Fig.~\ref{fig_vorticity}.
Blue denotes strongly elliptic regions with $\hat Q \leq -1$; Red denotes strongly hyperbolic regions with $\hat Q \geq 1$; and Yellow denotes intermediate regions with $-1<\hat Q<1$.}
\label{fig_weiss}
 \end{center}
\end{figure}

Figure~\ref{fig_weiss} shows a snapshot of the spatial distribution of the three-level, normalized Weiss field, $\hat{Q}=Q/Q_0$, in double periodic and bounded domains, corresponding to the flow field in Fig.~\ref{fig_vorticity}. The values of $Q_0$ are $170$ and $300$ for the periodic and confined domain, respectively. Consistent with the fact that elliptic regions are dominated by rotation, the negative regions are concentrated inside cyclones, $\omega>0$, and anti-cyclones, $\omega<0$, i.e.  the coherent vortices. The circulation cells surrounding  the vortices correspond to strongly hyperbolic regions which are dominated by deformation. The background turbulent field is characterized by small intermediate positive and negative values of the Weiss field. Consistent with the no-slip boundary conditions, the boundary in the circular domain exhibits high values of $|Q|$. 

\section{Residence time}

\begin{figure}[!htb]
 \begin{center}
 \includegraphics[scale=0.5]{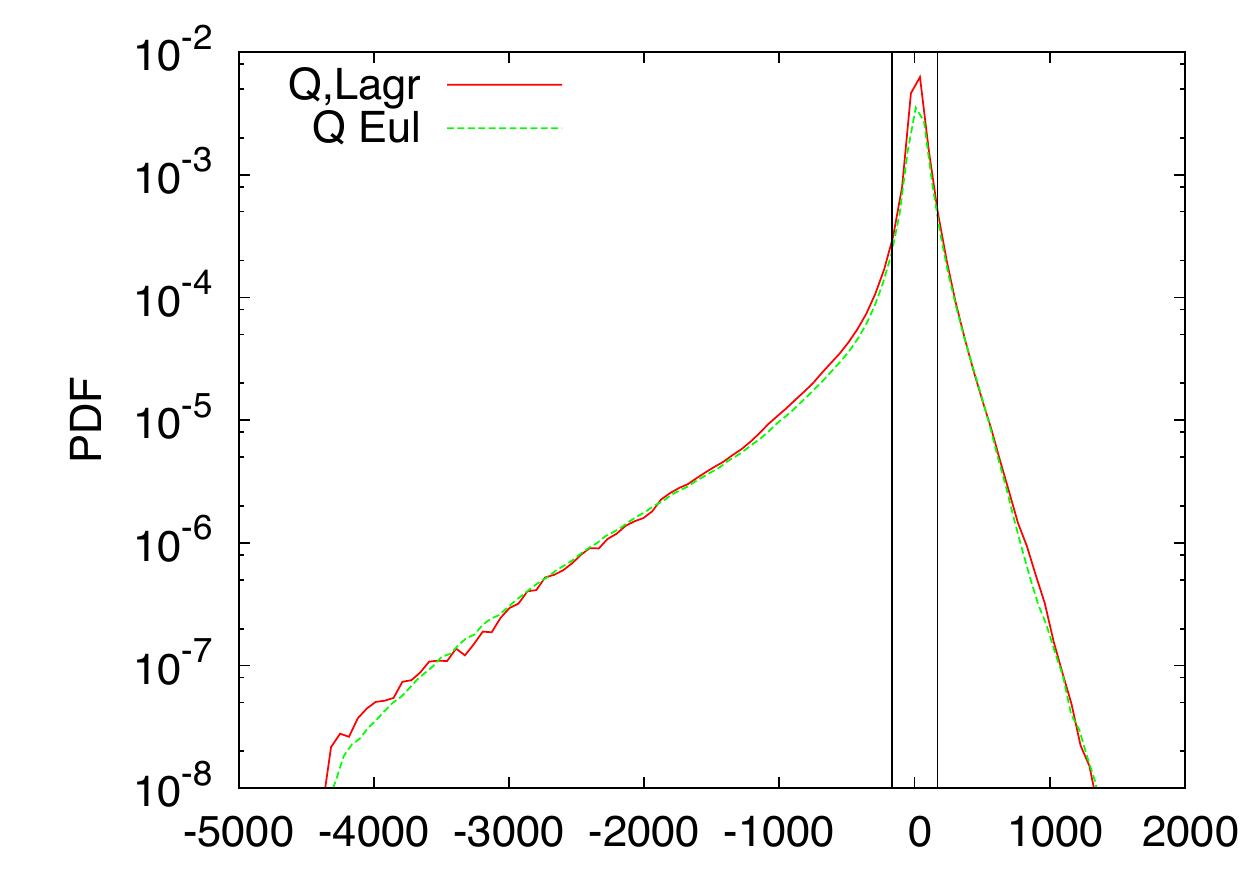}
 \includegraphics[scale=0.5]{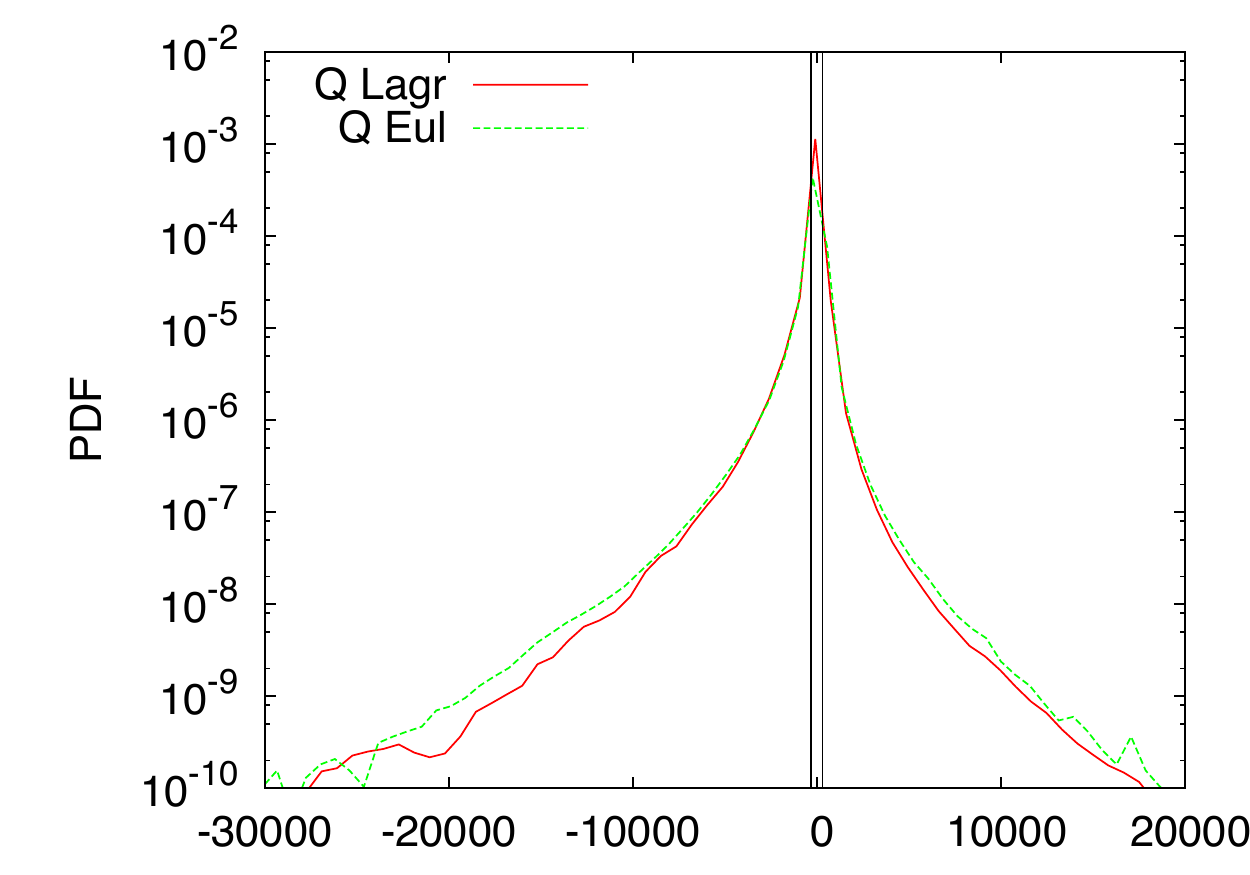}
\caption{
Probability density function of Lagrangian (red solid line) and Eulerian (green dashed line) Weiss field in two-dimensional forced turbulence in a 
periodic (top panel) and bounded (bottom panel) domain.}
\label{fig_pdfs_weiss}
 \end{center}
\end{figure}

The space-time dependent function $Q(x,y,t)$ in Eq.~(\ref{weiss}) 
is the Eulerian definition of the Weiss field.  Alternatively, given a trajectory $(x(t),y(t))$ obtained from the solution of 
Eq.~(\ref{hamilton}), the Lagrangian Weiss field is defined as the time dependent function $Q_L(t)=Q(x(t),y(t),t)$, i.e., the value of $Q$ along the particle trajectory. 
Figure~\ref{fig_pdfs_weiss} shows the probability density functions (pdfs) of the Eulerian Weiss field and the Lagrangian Weiss field for periodic and bounded domains. The Eulerian pdfs are obtained from a histogram of the spatial values of $Q$ at a fixed time, and the Lagrangian pdfs are obtained form the histogram of values along the trajectories of a large ensemble of particles.  The relatively good agreement of both pdfs indicates a consistent statistical sampling of the Lagrangian initial conditions.  As Figs.~\ref{fig_vorticity} and \ref{fig_weiss} show, there are typically more coherent vortices in the  periodic domain than in the bounded domain. This explains the asymmetry of the pdf in Fig.~\ref{fig_pdfs_weiss} towards negative $Q$ values in the periodic case. 
An interesting difference is also observed in the decay of the tails. In the bounded case, the pdf exhibits slowly decaying algebraic-type tails indicative of very large intermittent values of the Weiss field coming from the boundary. In contrast, in the periodic case the pdf decays exponentially. 

\begin{figure}[!htb]
 \begin{center}
 \includegraphics[scale=0.4]{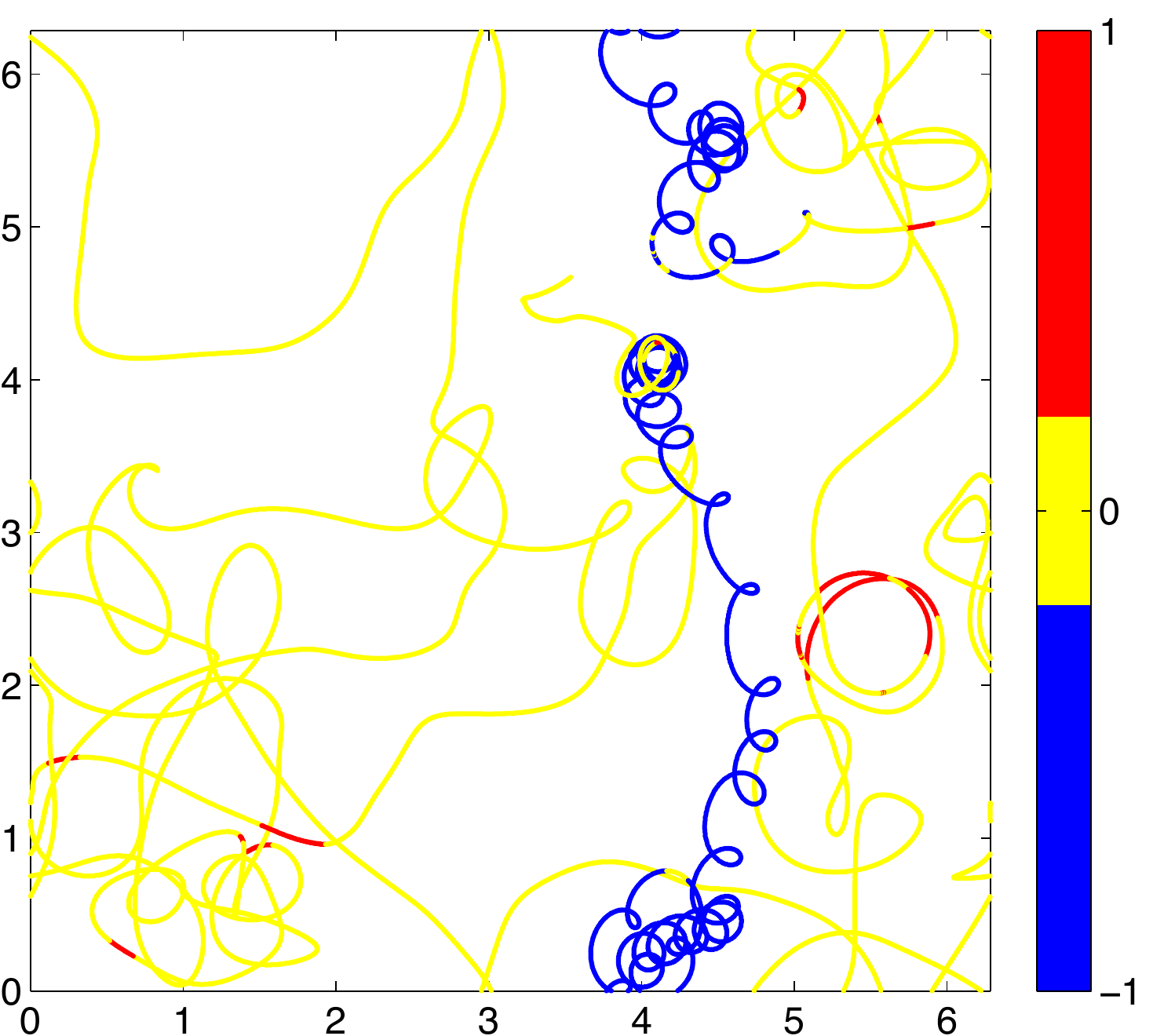}
 \includegraphics[scale=0.4]{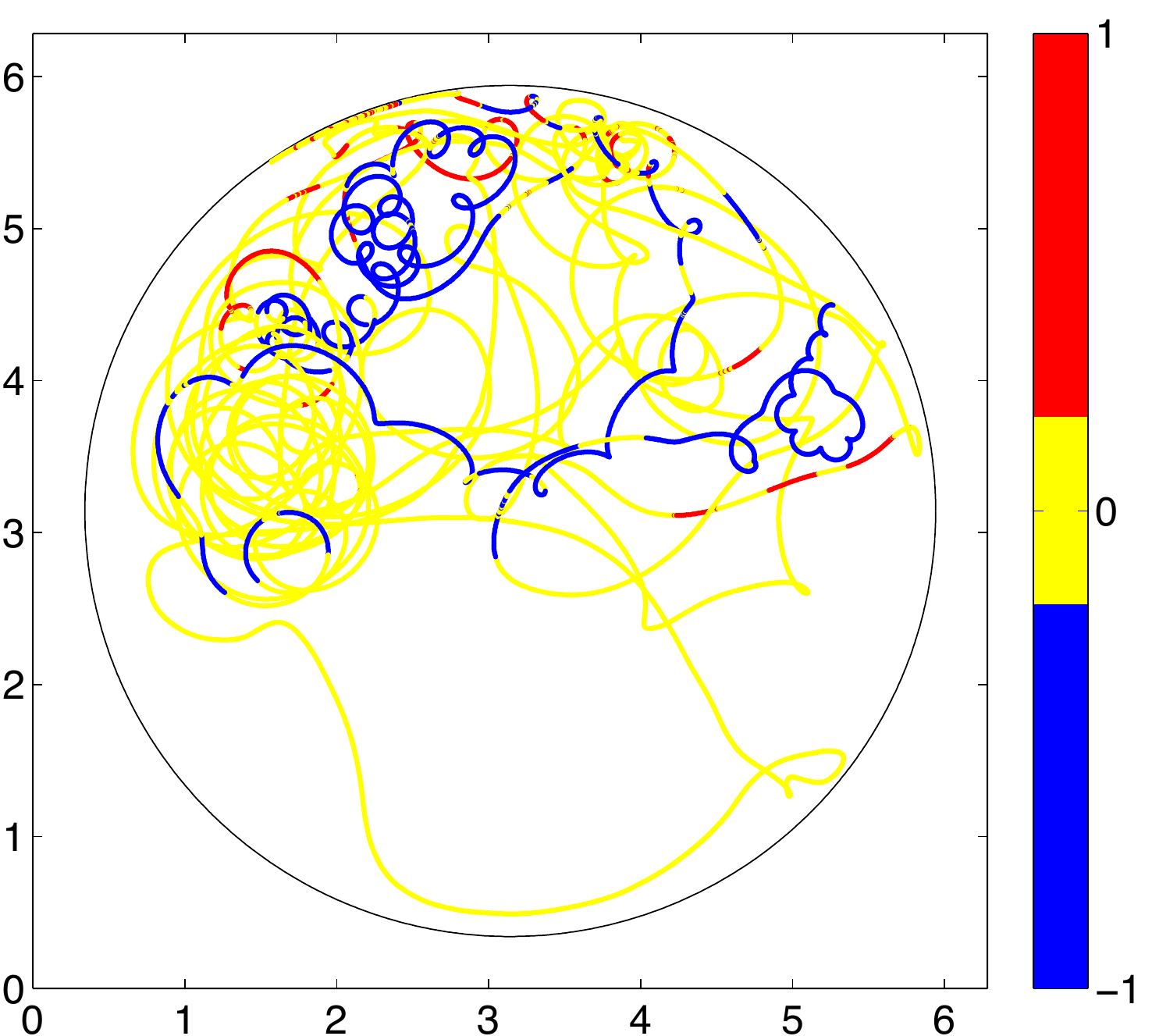}
\caption{
Typical passive tracer orbits in two-dimensional forced turbulence color-coded by the instantaneous value of the Lagrangian Weiss field in a periodic (top panel) and bounded (bottom panel) domain. 
 Blue denotes strongly elliptic regions with $\hat Q =-1$; Red denotes strongly hyperbolic regions with $\hat Q =1$; and Yellow denotes intermediate regions with $-1<\hat Q<1$.}
\label{fig_tracers_weiss}
 \end{center}
\end{figure}

Figure~\ref{fig_tracers_weiss} shows  typical passive tracer trajectories in the periodic and the bounded domains colored by the instantaneous value of the three-level Lagrangian Weiss field. The blue sections of the trajectories exhibit the expected spiraling motion resulting from the vortex trapping in  elliptic regions ($\hat Q <-1$) which move around the domain. The yellow section denotes the incursions of the particle throughout the turbulent background, and red denotes the sections of the orbit in the circulating cells surrounding the vortices.  One of the main thrusts of this paper is the use of the Lagrangian Weiss field as a method to characterize the statistics of passive tracers. 

\begin{figure}[!htb]
 \begin{center}
   \includegraphics[scale=0.5]{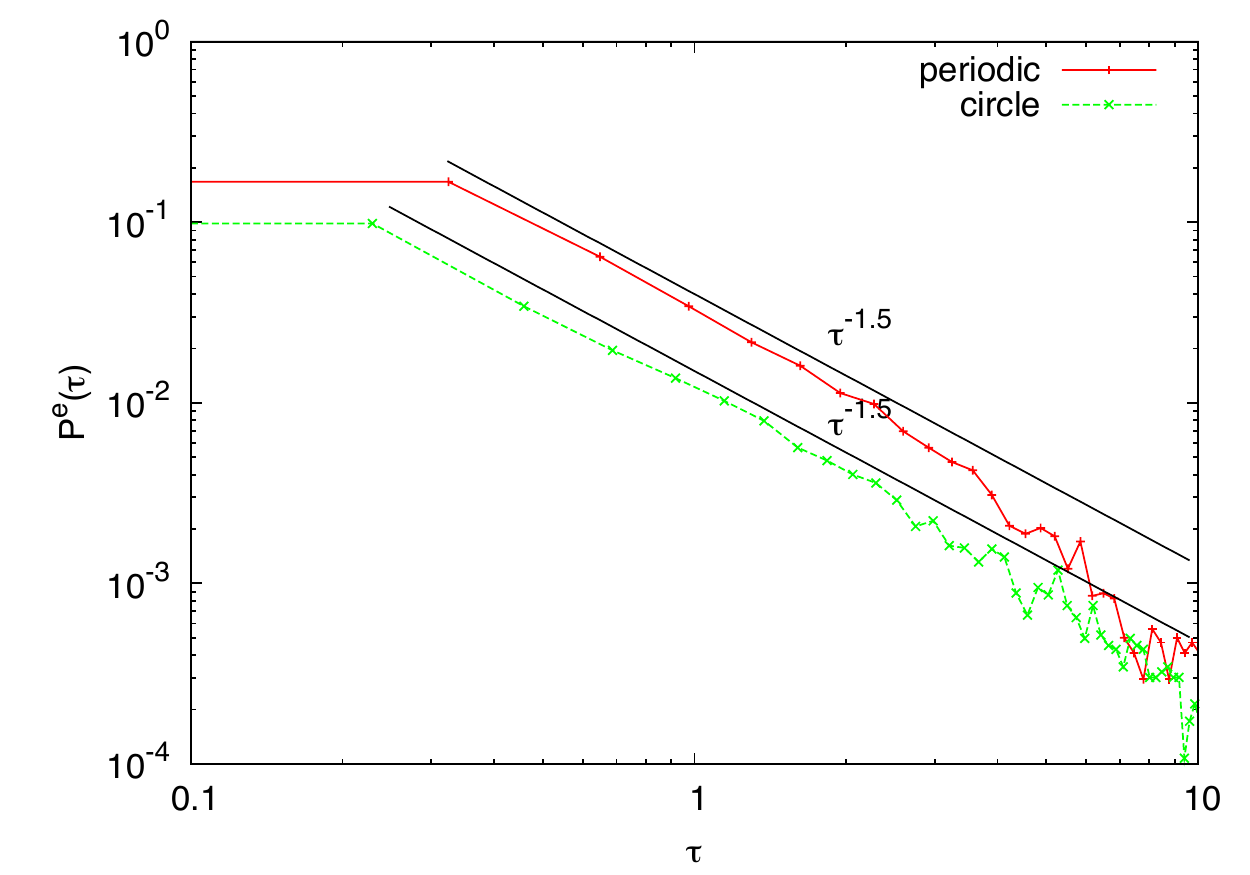}
   \includegraphics[scale=0.5]{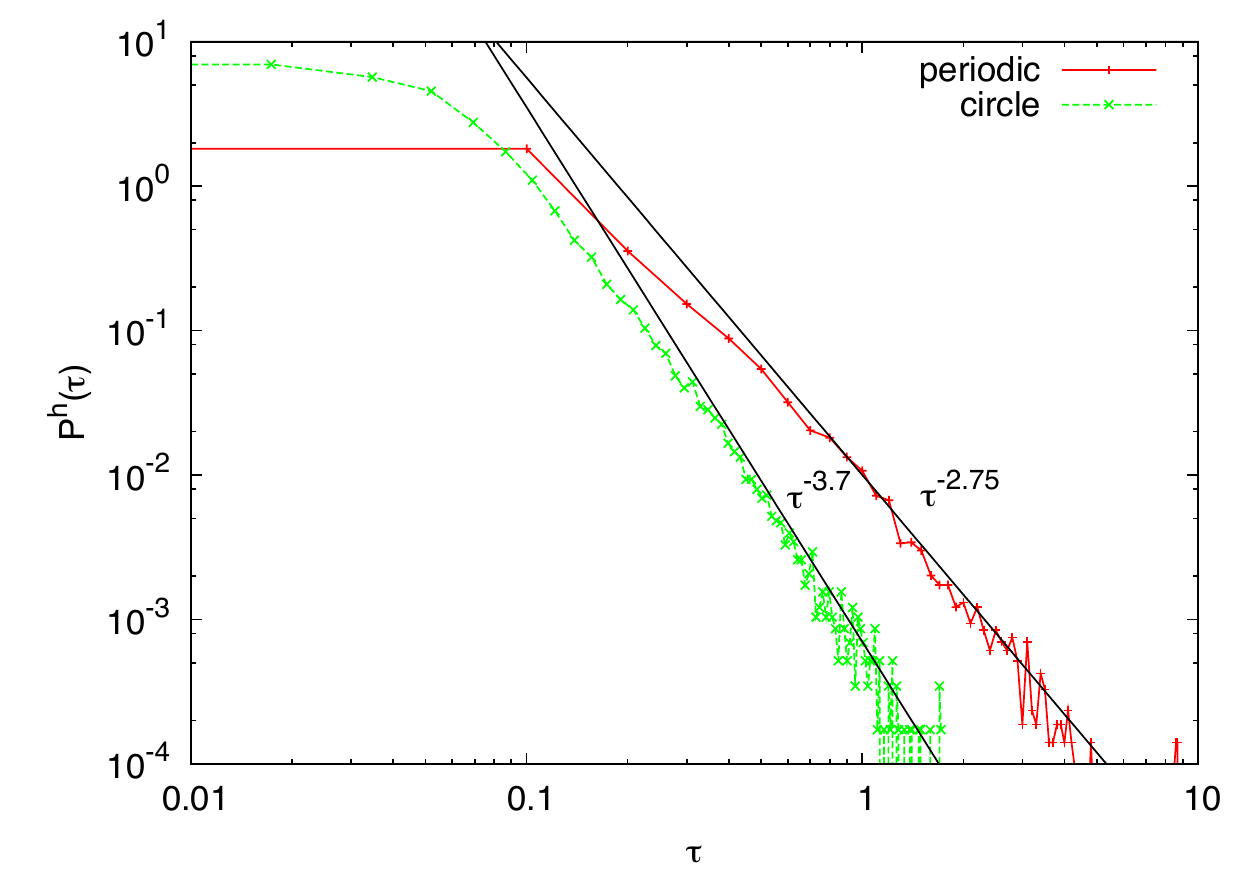}
\includegraphics[scale=0.5]{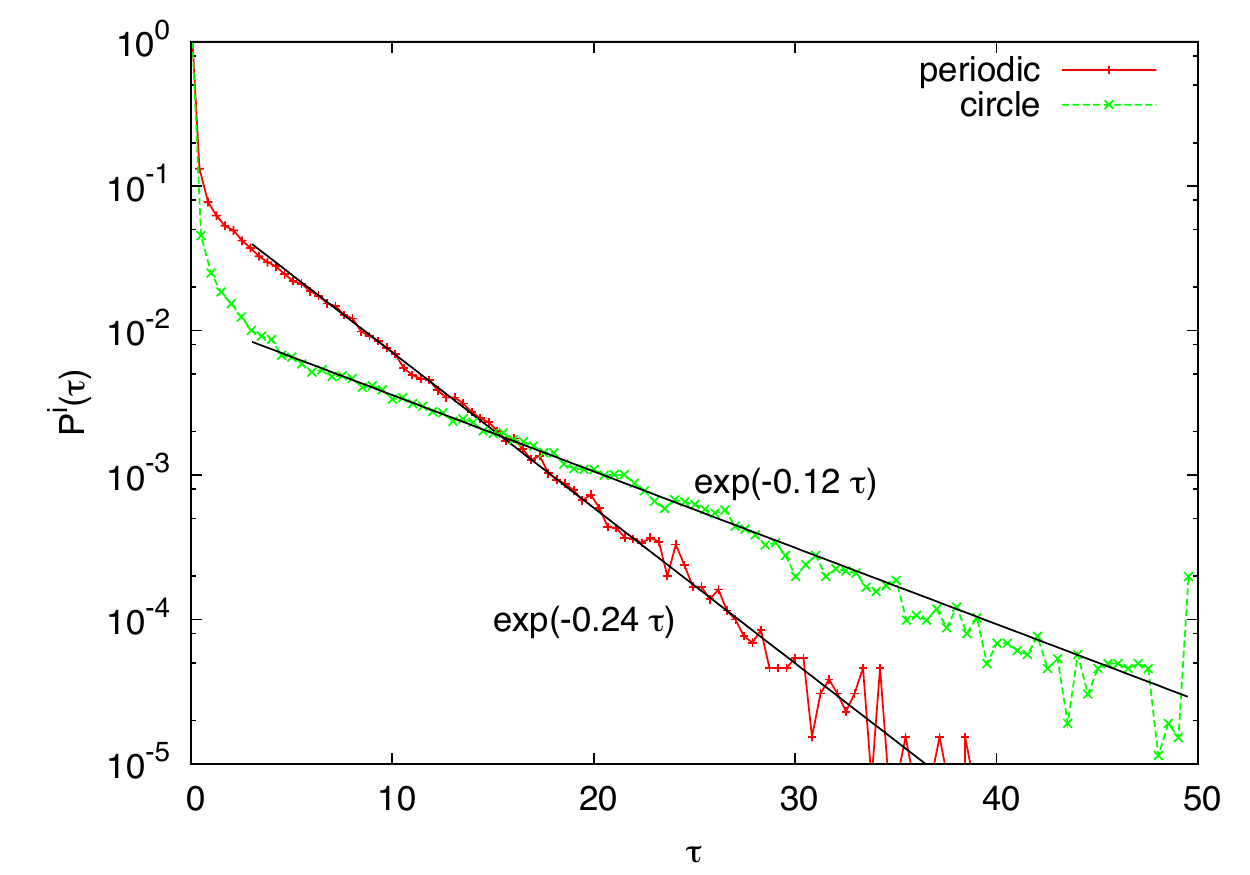}
\caption{Probability density functions of residence time for bounded (green) and periodic (red) domains.
Top: pdf, $P^e(\tau)$, in strongly elliptic, $ \hat{Q} <-1$, regions.
Middle: pdf, $P^h(\tau)$, in strongly hyperbolic, $\hat{Q} >1$, regions.
Bottom: pdf, $P^i(\tau)$, in intermediate, $-1<\hat{Q} <1$, regions.}
\label{pdf_residence_time_three_weiss}
 \end{center}
\end{figure}

In the double periodic and bounded domains, it is observed that  particles tend to spend relatively short times in the strongly hyperbolic regions. This is to be expected since these regions are unstable from a dynamical systems perspective. On the other hand, the relatively long stay of particles in strongly elliptic regions results from the trapping properties of vorticity dominated regions. To quantify these ideas we show in Fig.~\ref{pdf_residence_time_three_weiss} the pdfs of the residence time, $\tau$,  in the strongly elliptic, $P^e(\tau)$, strongly hyperbolic, $P^h(\tau)$, and intermediate, $P^i(\tau)$ regions. These pdfs determine the probability that a given Lagrangian tracer stays in a region with the same value of the three-level normalized Weiss field for a given time $\tau$. The Gaussian fluctuations characteristic of the turbulent background result in the exponential decay of $P^i(\tau)$, consistent with the results in Ref.~\cite{Moisy_2006} for three-dimensional turbulence. A possible explanation is that the particle in the intermediate zone can be related to a Poisson process which is characterized by a small time-correlation and an exponential pdf.  In contrast, $P^e(\tau)$ and $P^h(\tau)$ exhibit a self-similar behavior, corresponding to algebraic tails in the pdf. This is likely caused by the weak chaos, stickiness, and strong correlations characteristic of the vortex cores and the circulating cells surrounding the vortices. The strong dynamical instability of the circulating cells is responsible for the considerably larger decay exponent of $P^h(\tau)$ which is consistent with the relatively shorter red sections ($\hat Q>1$) of the particle orbits shown in Fig.~\ref{fig_tracers_weiss}. Because vortex trapping is a local phenomenon insensitive to boundaries, $P^e(\tau)$ exhibits the same decay in the periodic and the bounded domain. However, since particles in the turbulent background wander throughout the entire domain $P^i(\tau)$ shows some dependence on the boundary. 

The proposed use of the Lagrangian Weiss field  in the  definition of the residence time could be of value in the study of non-diffusive transport in fluids and plasmas. 
Early efforts in the study of  passive scalar transport  in flows exhibiting turbulence and chaotic advection were largely based on  
diffusion models. In these models, it is implicitly assumed that the particle trajectories can be described as an uncorrelated, Gaussian, Markovian stochastic process. However, several analytical, numerical, and experimental studies have pointed out the limitations of these statistical assumptions, and recent works have focused on the development of more general models,
see for example Ref.~ \cite{anomalous_transp} and references therein. A particularly productive approach is based on the use of continuous time random walk models, see for example Ref.~\cite{metzler} and references therein. These  models generalize the standard Brownian random walk by allowing more general, non-Gaussian (L\'evy) pdfs describing the particles' jumps and, most importantly, by incorporating waiting time distribution functions to account for particle trapping. 
The knowledge of the jump distribution and the waiting time distributions opens the possibility of constructing macroscopic models of non-diffusive transport using fractional diffusion operators. 
The characterization of waiting time distributions in particle transport studies in structured flows with simple spatio-temporal dependence is relatively simple, see e.g. Ref.~\cite{del_castillo_1998}. However, in turbulent flows the trapping structures exhibit random motion themselves  and can appear and disappear  in the course of the numerical calculation, e.g. 
Ref.~\cite{del_castillo_2005}.  When this is the case it is difficult to adopt an objective definition of a trapping event, and the construction of the waiting time pdf becomes nontrivial.  One possible approach is to use the pdf of residence time in strongly elliptic regions, $P^{e}(\tau)$, to construct the waiting time distribution.  This pdf provides an objective quantitative measure of the time a particle stays on a vortex even when, as shown in Fig.~\ref{fig_tracers_weiss}, the vortex moves. 

\begin{figure}[!htb]
 \begin{center}
   \includegraphics[scale=0.7]{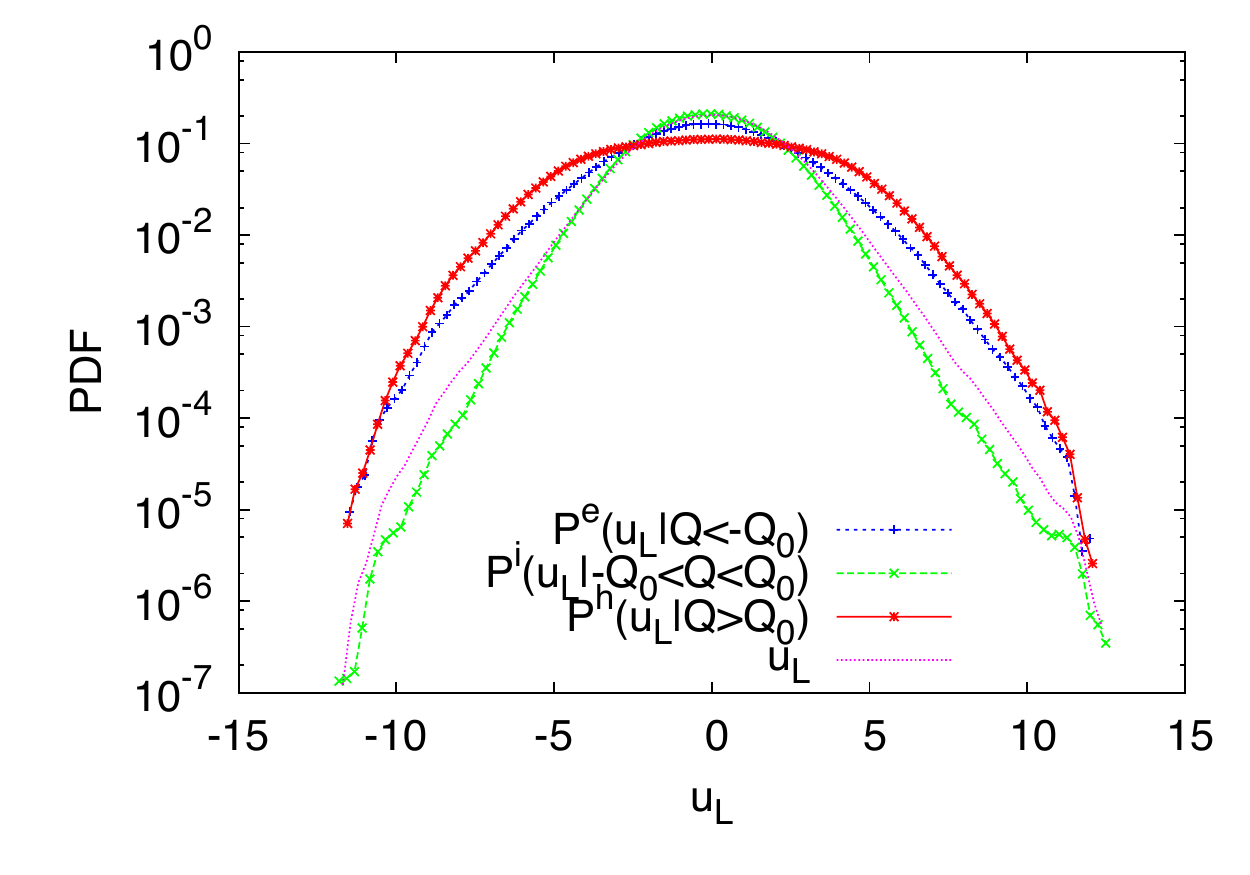}
   \includegraphics[scale=0.7]{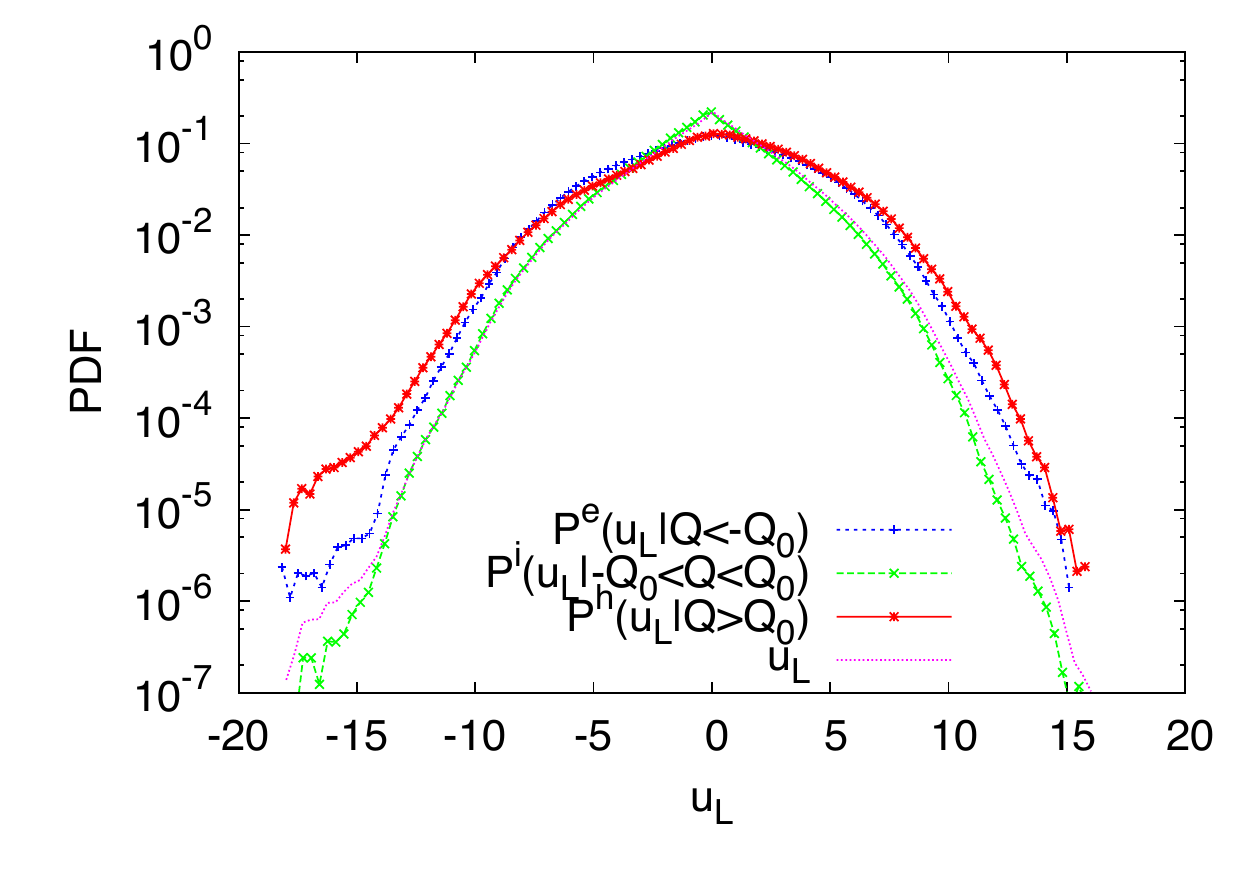}
\caption{Conditional pdf of Lagrangian velocity in the $x$-direction with respect to the 
three-level Weiss field  in the double periodic domain (top panel) and in the bounded circular domain (bottom panel).}
\label{pdf_velocity}
 \end{center}
\end{figure}

\begin{figure}[!htb]
 \begin{center}
   \includegraphics[scale=0.7]{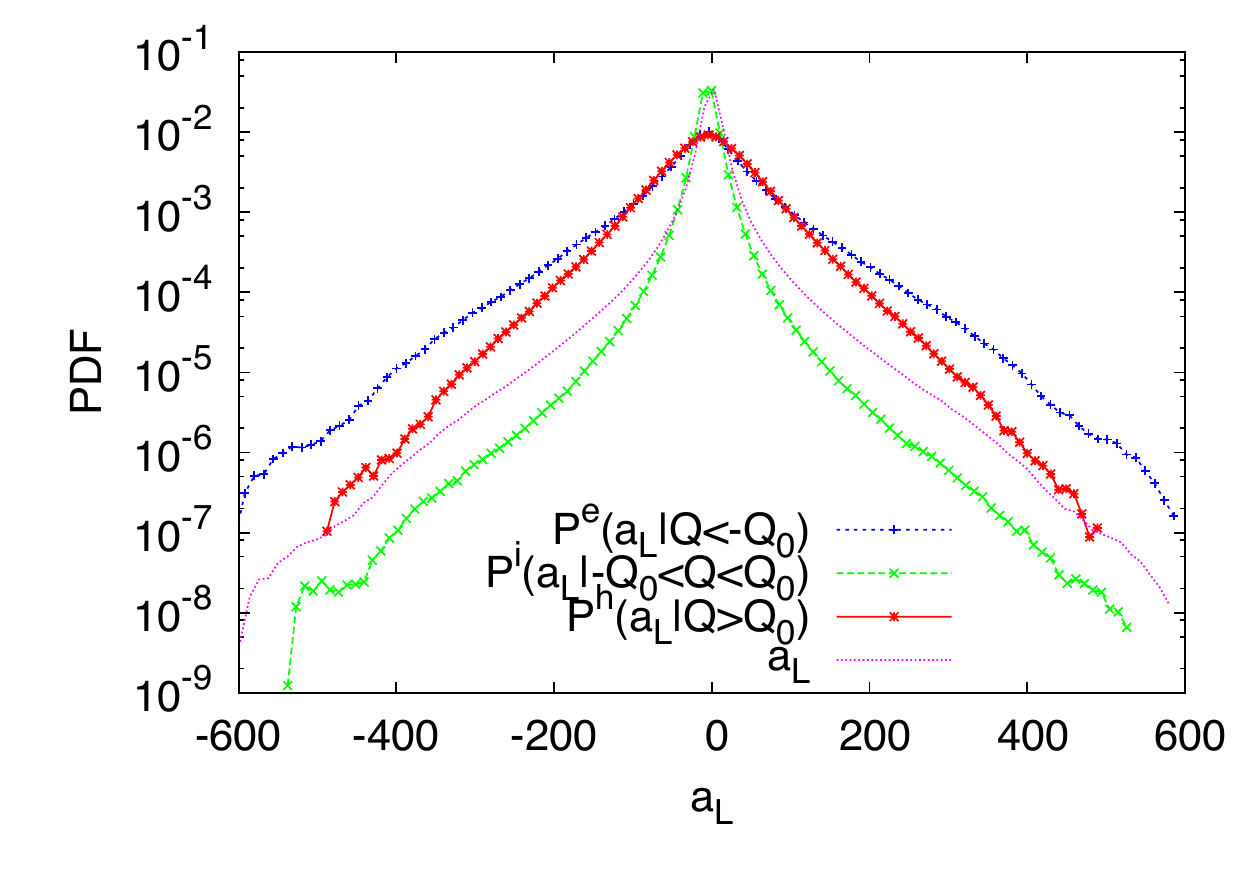}
   \includegraphics[scale=0.7]{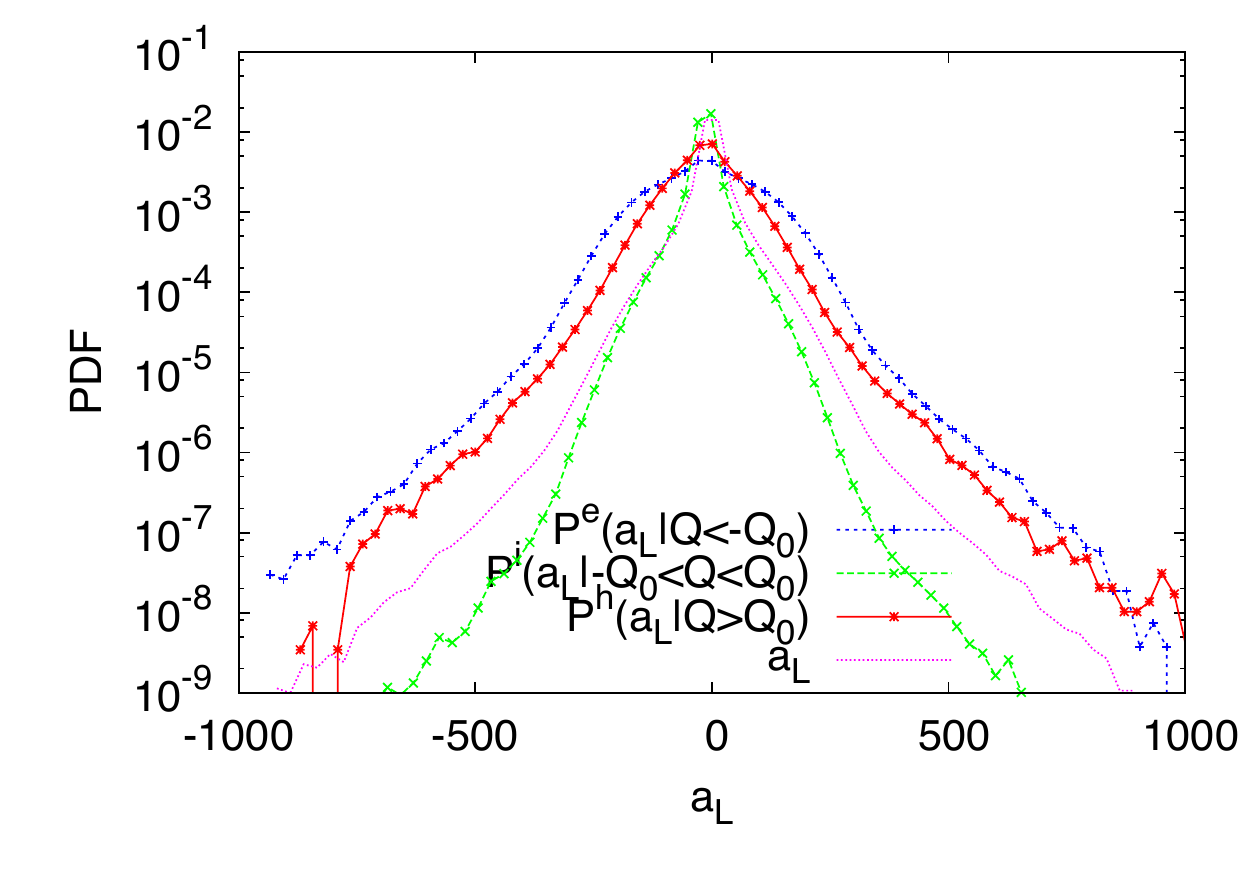}
\caption{Conditional pdf of Lagrangian acceleration in the $x$-direction with respect to the 
three-level Weiss field  in the double periodic domain (top panel) and in the bounded circular domain (bottom panel).}
\label{pdf_acceleration}
 \end{center}
\end{figure}

\begin{figure}[!htb]
 \begin{center}
   \includegraphics[scale=0.7]{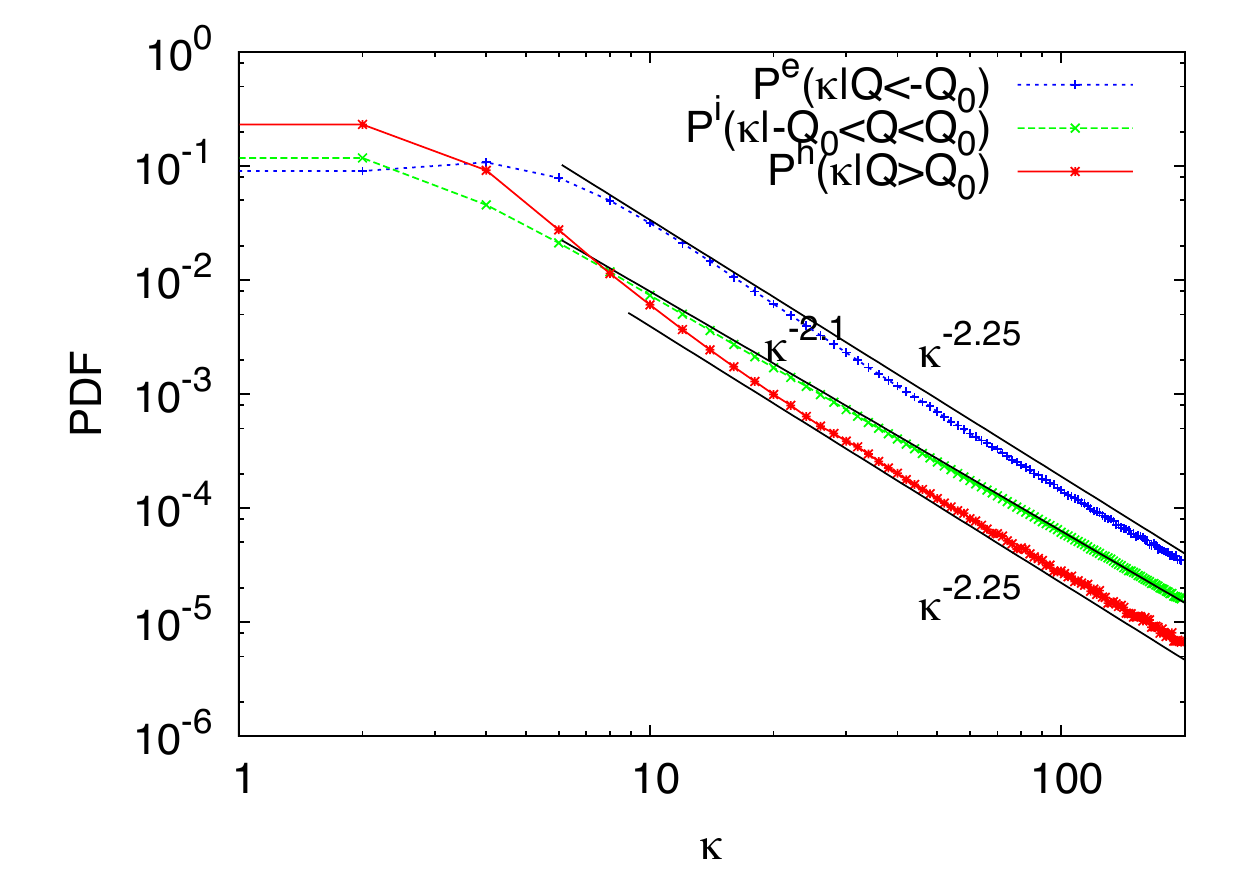}
   \includegraphics[scale=0.7]{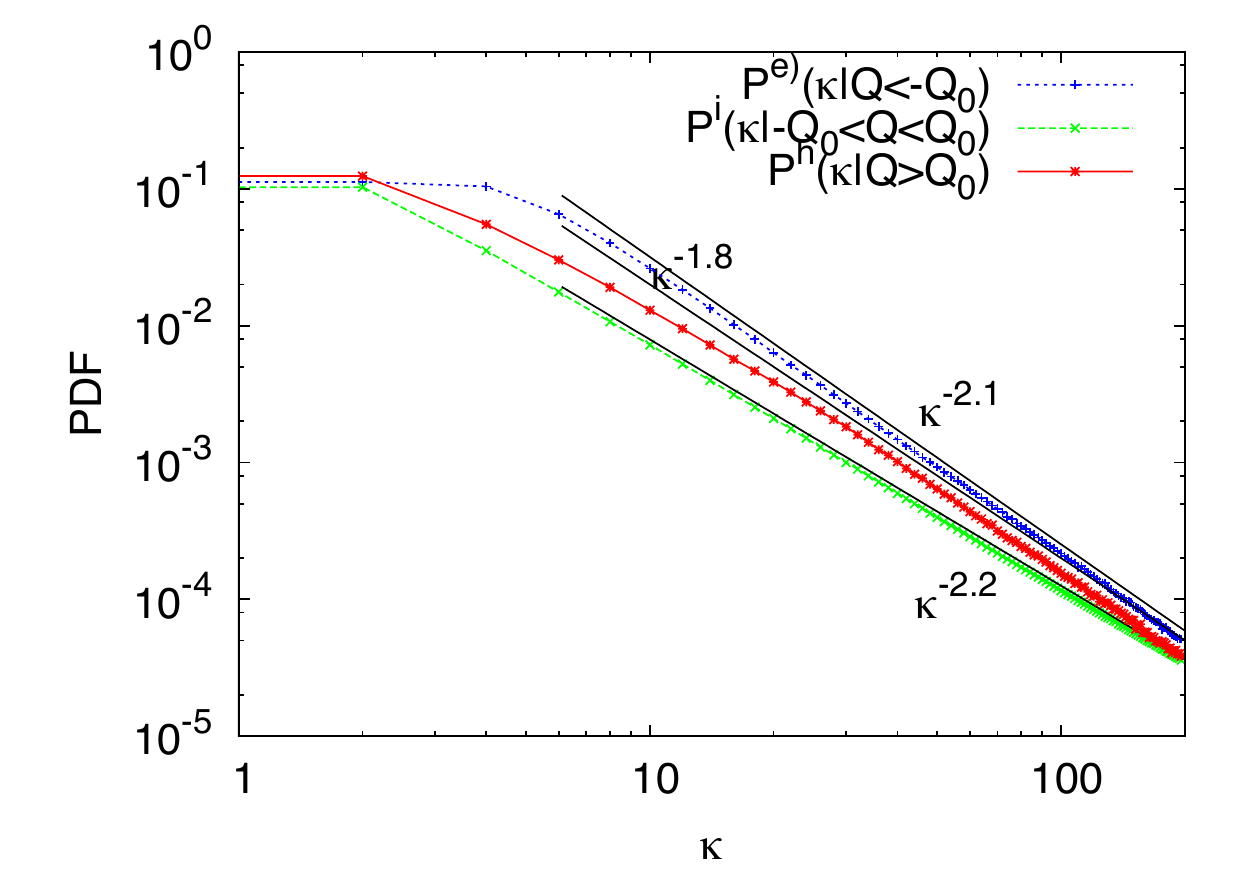}
\caption{Conditional pdf of curvature with respect to the three-level Weiss field
in the double periodic domain (top panel) and in the bounded circular domain (bottom panel).}
\label{pdf_curvature}
 \end{center}
\end{figure}

\section{Conditional Lagrangian statistics}

 In this section we study the dependence of the statistics of the Lagrangian velocity, acceleration, and curvature on the flow topology, using the  conditional pdfs of these quantities on the values of the Weiss field. For each field (velocity, acceleration, and curvature) we construct three conditional pdfs determining the probability of a specific value provided the particle is in a strongly elliptic, strongly hyperbolic, or an intermediate region. For example,  $P^{e}(a_L | Q \leq -Q_0) d a_L$ gives the probability    that the Lagrangian acceleration has a value in the range $(a_L, a_L+d a_L)$ while the particle is in a strongly elliptic region. The definition of the conditional probabilities for the other fields (velocity and curvature) and for the other regions of the flow (intermediate and strongly hyperbolic) is straightforward. 

Figure~\ref{pdf_velocity} shows the conditional pdfs of the Lagrangian velocity in the $x$-direction in periodic and bounded domains.  In the periodic case, $P^{e}$ and $P^h$ have Gaussian-type dependence, i.e., in log-normal scale they exhibit parabolic profiles. On the other hand, $P^i$  has an exponential-type dependence, as shown by 
the linear decaying tails in  the log-normal scale. This indicates that in the  intermediate region comprising the turbulent background, the velocity is more intermittent than inside the vortices and in the straining regions surrounding them. 
However, as the bottom panel in Fig.~\ref{pdf_velocity} shows, this distinction is less clear in the bounded domain. The corresponding pdfs in the $y$-direction (not shown) have very similar behavior. 

The conditional pdfs of the Lagrangian acceleration in the $x$-direction are shown in Fig.~\ref{pdf_acceleration}. 
It is observed that the level of intermittency (characterized by the heavy tails) is comparable  in the double periodic and bounded domains. In particular, in all cases the tails exhibit exponential decay. This is in contrast to what is observed in freely decaying two-dimensional turbulence \cite{Kadoch_PRL_2008} in which the circular domain shows a significant higher level of intermittency. 
The conditional pdf for the intermediate values of the Weiss field, $P^{i}(a_L|-Q_0<Q<Q_0)$, shows a sharp peak around zero indicating a stronger intermittency of the acceleration in the intermediate region. 

Figure~\ref{pdf_curvature} shows the pdfs of the conditional pdf of the Lagrangian curvature. In all cases self-similar power law behavior is observed with a decay exponent of order $-2$. As discussed in Ref.~\cite{xu_PRL_2007}, this can be explained 
as follows. In isotropic turbulence, the Lagrangian velocity $u_L$ is close to a Gaussian distribution, which implies that $u_L^2$ is distributed accordingly to a chi-squared distribution and thus $1/{u_L}^2$ is distributed according to an inverse chi-squared distribution whose tails decay as $(1/{u_L}^2)^{-2}$. Since $\kappa=|a_n| {u_L}^{-2}$, this implies that, under the isotropic turbulence and Gaussian assumptions, the tails of the pdf of $\kappa$ should decay as $\kappa^{-2}$. Note that in three dimensions the same argument yields the exponent $-2.5$ \cite{xu_PRL_2007}. The slight departures from this simple scaling observed in the numerical is a consequence of  the nontrivial statistics and the dependence of the curvature on $a_n$. 

\section{Summary and conclusions}

In this paper we have presented a study of the relationship between Lagrangian statistics  and flow topology which we characterized using the Weiss criterion. The Weiss criterion provides a simplified tool to partition the flow field into topologically different regions. 
By flow topology we mean the topology of the streamlines. Since the flow is assumed two-dimensional and incompressible the streamlines corresponds to the level sets of the  Hamiltonian describing the Lagrangian orbits. 
Most of the previous studies have considered the  Weiss criterion in the Eulerian context. Here we focused on the Lagrangian Weiss field. That is, we considered the Weiss field along individual particle trajectories rather than  as a function of the spatial coordinates in the flow.  The flow field corresponded to forced two-dimensional Navier-Stokes turbulence in double periodic and circularly bounded domains. The study of the statistics focused on the residence time and in the conditional probability density functions of the Lagrangian velocity, acceleration, and curvature. 

The flow topology was partitioned into strongly elliptic regions for which $Q \leq -Q_0$, strongly hyperbolic regions for which
$Q \geq Q_0$, and intermediate regions for which $-Q_0 < Q  < Q_0$, where $Q_0=\sqrt{\langle Q^2 \rangle}$, is the standard deviation of $Q$ values. In terms of the three levels Weiss field, these regions correspond to $\hat Q \leq -1$,
$\hat Q \geq 1$, and $-1<\hat Q <1$.  The strongly elliptic regions correspond to the vortices, the strongly hyperbolic regions correspond to the large strain regions surrounding the vortices, and the intermediate regions to the turbulence background.

A very good agreement was observed between the pdf of the Eulerian Weiss field and the pdf of the Lagrangian Weiss field, confirming statistical convergence. In the double periodic domain, the pdf of the Weiss field exhibited a negative skewness consistent with the fact that in periodic domains the flow is dominated by coherent vortex structures, i.e., elliptic regions. On the other hand, in the circular domain, the elliptic and hyperbolic regions seem to be statistically similar.
The pdf of the Weiss field in the bounded domain is more intermittent than in the periodic domain  due to the large vorticity and deformation fluctuations at the boundary resulting from the non-slip condition. 

To characterize statistically the time that particles spend in the topologically different regions of the space, we computed the  pdfs of the residence time, $\tau$, in the strongly elliptic regions, $P^{e}(\tau)$, in the strongly hyperbolic regions, $P^{h}(\tau)$, and in the intermediate regions, $P^{i}(\tau)$. The pdfs $P^{e}(\tau)$ and $P^{h}(\tau)$ exhibit algebraic decaying tails, and the pdf $P^{i}(\tau)$ exhibits exponential decaying tails. $P^{e}(\tau)$ provides an objective quantitative measures of the time particles stays on a vortex even though,  the vortex can exhibit a complicated motion. This motivates the use of this pdf to construct waiting time pdfs in continuous time random walk descriptions of anomalous diffusion in turbulent systems with coherent trapping structures. 

To study the dependence of the Lagrangian statistics on the flow topology, we investigated the conditional pdfs of the Lagrangian velocity, acceleration, and curvature on the values of the Lagrangian Weiss field. The conditional pdfs of the Lagrangian velocity  typically have Gaussian behavior in the periodic and in the bounded  domain, with the exception of the conditional pdf on the intermediate values of the Weiss field in the periodic domain that shows exponential decay. 
The conditional pdfs of the Lagrangian acceleration have a comparable level of intermittency in the periodic and the bounded domains. This is in contrast to the freely decaying turbulence case, for which the circular domain exhibits a significant higher level of intermmitency.  
Finally, the conditional pdfs of the Lagrangian curvature displayed in all cases self-similar power law behavior with a decay exponent of order $-2$, reminiscent of the Gaussian character of the single point velocity statistics.

\subsection*{Acknowledgments}
DCN acknowledge support  from the Oak Ridge National Laboratory, managed by UT-Battelle, LLC, for the U.S. Department of Energy under contract DE-AC05-00OR22725. DCN also gratefully acknowledges the support and hospitality of the \'Ecole Centrale de Marseille for the visiting positions during the elaboration of this work. BK and WB thank F. Moisy for fruitful discussions in the framework of the GDR turbulence, and G. Keetels is acknowledged for discussions on the forcing term.

\end{document}